\documentclass[twocolumn,showpacs,preprintnumbers,amsmath,amssymb]{revtex4}

\usepackage{graphicx}
\usepackage{dcolumn}
\usepackage{bm}

\begin{document}

\title{Local Potential Model of the Hoyle Band in $^{12}$C }

\author{B. Buck$^{\ast}$, A.C. Merchant$^{\ast}$ and S.M. Perez$^{\dagger\#}$}

\affiliation{%
$\ast$ Department of Physics, University of Oxford, Theoretical Physics, 
1 Keble Road, Oxford OX1 3NP, UK. \\
$\dagger$ 
Department of Physics, University of Cape Town, Private Bag, 
Rondebosch 7700, South Africa. \\
\# iThemba LABS, P.O.Box 722, Somerset West 7129, South Africa.
}%
\date{\today}
\begin{abstract}

\noindent We describe the excited $0^+$ state of $^{12}$C at 7.654
MeV, often called the Hoyle state, in terms of a local potential
$^8$Be+$\alpha$ cluster model. We use a previously published 
prescription for the cluster-core potential to solve the Schr\"odinger 
equation to obtain wave functions for this state, and also for 
higher angular momentum states of the same system. We calculate energies, 
widths and charge radii for the resulting band of states, with
particular emphasis on the recently discovered $2^+$ state. We 
examine various choices of the global quantum number $G=2n+L$ 
for the cluster-core relative motion, and find that $G=6$ leads 
to the most coherent description of the properties of the states 
and is consistent with recent experimental data on the $L=2$ state.

\end{abstract}
\maketitle

\section{Introduction} 

When considering how carbon could be produced in stellar
nucleosynthesis, Hoyle \cite{[Hoyle_1]} famously predicted
that the $^{12}$C nucleus must have an excited $0^+$ state 
in the vicinity of the 3-$\alpha$ breakup threshold. Such a state 
was duly found \cite{[Dunbar]} in short order on the basis of 
Hoyle's suggestion. Its existence was essential for $^{12}$C to be 
produced at an adequate rate to open up the pathways to the synthesis 
of still heavier nuclei \cite{[Hoyle_2]}. Not only had 
this state escaped both experimental detection and theoretical 
prediction up to that point, but it has continued to pose severe 
challenges to nuclear structure models ever since.

The shell model struggles to describe any low-lying excited $0^+$
states in p-shell and sd-shell nuclei. To this day, the most 
advanced no-core shell model has not succeeded in reproducing
the excitation energy of the Hoyle state \cite{[Navratil]}. Models invoking a 
3$\alpha$ chain \cite{[Morinaga],[Horiuchi_1],[Me_Rae]} 
were able to accommodate the excitation energy, but could not 
reproduce the decay width of the Hoyle state. Better success was 
enjoyed by orthogonality condition model calculations of the 
3$\alpha$ system, using a semi-microscopic $\alpha-\alpha$ interaction 
\cite{[Horiuchi_2],[Saito]}, which gave a suitably sized $\alpha$-decay 
width in the $^8$Be$(0_1^+) + \alpha$ channel. This suggested that the
dominant structure of the Hoyle state must be $^8$Be$(0_1^+) + \alpha$
in a relative s-state. Such a conclusion was backed up by fully
microscopic $3\alpha$ calculations using the Resonating Group
and Generator Coordinate Methods \cite{[Hack],[Kamimura],[Uegaki],[Descouv]}. 
More recently fully microscopic calculations using Antisymmetrized Molecular
Dynamics (AMD) \cite{[Enyo]} and Fermionic Molecular dynamics (FMD) \cite{[Feld]}
have been able to give a good account of the low-lying spectrum of 
$^{12}$C without assuming alpha clustering {\it a priori\/}. For the
particular case of the very loosely bound Hoyle state a three-$\alpha$
condensate wave function \cite{[FunakiCondensate]} has also been shown to have
a large overlap with the FMD wave function. The current situation is well
summarized in a recent review \cite{[HIK]}.

In parallel with these theoretical developments, new life has recently
been breathed into the experimental program. Given that the Hoyle 
state has a large mean square radius, it is natural to suggest 
calculating higher angular momentum states with this same structure
thereby producing a band of excited states with a large moment of 
inertia. Although Friedrich et al. \cite{[Friedrich]} claim 
there are no such rotational states,
most $3\alpha$ models do support them, with the $2^+$ state
generally expected to have an excitation energy in the region of 10 MeV.
Initial searches for this $2^+$ state via beta decay \cite{[Fynbo]} were
discouraging, but in recent times three separate experiments have 
provided evidence of its existence. Itoh et al. \cite{[Itoh]} see a $2^+$
state at 9.9 MeV, with a width of 1.0 MeV, via inelastic $\alpha$-particle
scattering in their $^{12}$C($\alpha,\alpha^{\prime}$) measurements. Freer
et al. \cite{[Freer]} see a $2^+$ state at 9.6 MeV, with a width of 0.6 MeV,
in their inelastic proton scattering $^{12}$C(p,p$^{\prime}$)
work. Gai \cite{[Gai]} confirms the existence of a $2^+$ state 
around 10 MeV (but without being able to measure a width) in 
photonuclear disintegration $^{12}$C($\gamma, 3\alpha$) studies.

In view of this resurgent interest in an excited Hoyle band we ask 
in this paper to what extent the known data can be accounted for (and further
excited states predicted) by a local potential $^8$Be+$\alpha$ cluster 
model. This model provides a physically transparent and calculationally
straightforward description of the energies, widths and charge radii
of the known states in the proposed Hoyle band. It also throws some light
on the question of how high in angular momentum such a band might
continue. In this way it can be a useful guide to experimental groups
searching for as yet unidentified higher $L$ states. It is also
illuminating to see how far one can get with a simple but physically 
motivated model. Although ideally the full rigours of a more
microscopic and computationally intensive approach might seem preferable,
a rather simplified effective nucleon-nucleon potential is needed to
carry this to fruition. It might be that a phenomenological
approach can produce a band of states with properties closer to the 
experimental values. In any
event it is interesting to compare the results from different 
theoretical treatments with each other as well as with experiment.

In the next section we describe our local potential $^8$Be+$\alpha$ cluster 
model. After that, we compare the model's results with the available data and 
discuss the possible existence of additional states. Finally, we summarize
our conclusions.

\section{Local potential cluster model}

The cluster model employed here was first proposed to study the
excited 4p-4h band of $^{16}$O (bandhead at 6.05 MeV) and the
ground state band of $^{20}$Ne \cite{[BDV]} and has
subsequently been applied to a wide range of nuclei across the 
Periodic Table from $^6$Li to $^{242}$Cm. However, it has not
previously been used for $^{12}$C, because the ground state band of
that nucleus is known to be oblate (from the sign of its quadrupole
moment \cite{[Vermeer]}) and a two-body cluster-core system inevitably 
produces a prolate deformation. 
This is because, for spinless constituents like $^8$Be and $\alpha$,
the quadrupole operator for the system reduces to
\begin{equation}
M(E2) = {(Z_1A_2^2 + Z_2A_1^2) \over (A_1 + A_2)^2 } R^2 Y_2(\hat{\bf R})
\end{equation}
The expectation value of this operator always yields a positive quadrupole
moment, indicating a prolate spheroidal shape. To obtain negative
quadrupole moments, appropriate to oblate shapes, it is necessary to
consider three-body systems. This outcome provides a large part of the
motivation behind attempts to model the ground state band of $^{12}$C
(and indeed the excited $3^-$ state at 9.64 MeV) as an equilateral
triangular arrangement of three $\alpha$ particles
\cite{[HafstadTeller]}.
However, the indications that the 
Hoyle state, and any associated rotations of it, have a large 
$^8$Be$(0_1^+) + \alpha$ component, mark out the Hoyle band as ideal 
territory for our local potential two-body cluster model.
Unfortunately, this does mean that we are unable to address any states of
the $^{12}$C system that do not have this particular structure. In
particular, we cannot model the ground state band and the excited
$3^-$ state mentioned above.

In general, we model a nucleus as two even-even sub-nuclei of mass 
$A_1$ and $A_2$ separated by a distance ${\bf R}$, interacting through 
a deep, local nuclear potential $V_N(R)$ and a Coulomb potential $V_C(R)$ 
appropriate to a point cluster and a uniformly charged spherical core.
The nuclear part has previously been parametrised in the form 
\cite{[univ]}:
\begin{eqnarray}
V_N(R) &=& -V_0 \Biggl \{ {x \over [1 + \exp{((R-R_0)/a))} ] } 
\nonumber \\\ &+&
{1-x \over [1+\exp{((R-R_0)/3a)} ]^3 } \Biggr \}
\end{eqnarray}
with parameter values given by
\begin{equation}
v_0 = 54.0 \ {\rm MeV,} \ \ \ a=0.73 \ {\rm fm,} \ \ \ {\rm and} 
\ \ \ x=0.33
\end{equation}
and $V_0$ related to $v_0$ by
\begin{equation}
V_0 ={\displaystyle  { A_1A_2 \over (A_1+A_2-1) } v_0 \over
  {\displaystyle { x \over [1 + \exp{(-R_0/a)} ] } +{1-x \over
      [1+\exp{(-R_0/3a)} ]^3 } } }.
\end{equation}
The radius parameter $R_0$ of the potential is determined by fitting
to the experimental energy of the spectrum's band head, and is linked
to the choice of $G=2n+L$ (see below). This potential is now incorporated 
into the cluster-core relative motion Hamiltonian $H_0({\bf R})$, 
and the resulting Schr\"odinger equation
\begin{eqnarray} 
H_0({\bf R}) \Phi_{GnL}({\bf R}) &=& E_{GnL}\Phi_{GnL}({\bf R}) 
\nonumber \\\ &=& 
E_{GnL}{ u_{GnL}(R) \over R}Y_{LM}(\theta, \phi). 
\end{eqnarray}
is solved numerically. We label the wave functions and energies with the 
global quantum number $G=2n+L$, where $n$ is the number of internal 
nodes in the radial wave function and $L$ the orbital angular momentum.

We must choose the value of $G$ large enough to guarantee that the  
Pauli exclusion principle is satisfied by excluding the cluster  
constituents from states occupied by the core nucleons. For the $^8$Be +
$\alpha$ system this requires $G \geq 4$. The appropriate choice of $G$
is not clear cut when we are not using an oscillator potential, although
we can certainly use oscillator considerations as a guide. The lowest 
allowed value, $G=4$, would correspond to packing the $\alpha$ nucleons 
into the p-shell, which seems an unlikely description of an excited state 
in $^{12}$C. Nevertheless, for completeness, we present calculations 
employing $G = 4, 6$ and 8, with a view to deciding {\it a posteriori\/}
on the basis of a comparison with experimental data which is the best
choice.
Indeed, only {\it a posteriori\/} can we say that our model, with any
eventually preferred value of $G$, is appropriate for describing the
Hoyle state and its associated band by comparing 
its predictions with the measured properties of those
states. 

Although numerical solution of the Schr\"odinger equation
without any restrictions on $G$ certainly does produce lower lying states,
they are Pauli forbidden and do not correspond to physical states of
$^{12}$C. Thus, we have no $^{12}$C bound states in our model and
cannot use it to describe the observed ground $0^+$ or excited $2^+$
(4.44 MeV) states of the nucleus. Equally, we cannot describe the
$3^-$ (9.64 MeV) state of $^{12}$C (as explained earlier). It is
possible to produce negative parity states in our model by solving the
Schr\"odinger equation with an odd value of $G$, but the resulting
energies are 10--20 MeV above the Hoyle state, and not of immediate
interest. Similarly, we could obtain negative parity states of $^{12}$C
by using even $G$ values in conjunction with an excited negative
parity state of the $^8$Be core. Again, the resulting excitation
energies in $^{12}$C are too high to be of interest in the current study.

As a further model extension we could include excitations of the
$^8$Be core into its $2^+$ and $4^+$ states. We do not do this here
because it would involve the introduction of more adjustable
parameters to describe the non-central interactions that accompany
such excitations, and we are trying to keep the number of
fitted quantities to a minimum. Our previous experience of
including core excitations in treatments of $^{16}$O \cite{[BBR_O16]} 
and $^{24}$Mg \cite{[BHM_24Mg]}
leads us to expect that they would not have a major effect on our
conclusions in the the $^8$Be$ + \alpha$ case. 

Solving the Schr\"odinger equation produces excitation energies and their
associated widths directly. The resulting wave functions can also be used 
to calculate mean square charge radii of the states from
\begin{eqnarray}
(Z_1 + Z_2) \langle R^2 (^{12}{\rm C}) \rangle &=& Z_1 \langle R^2 (^8{\rm Be})
\rangle + Z_2 \langle R^2 (\alpha) \rangle \nonumber \\\
&+& {(Z_1A_2^2 + Z_2A_1^2) \over (A_1 + A_2)^2 } \langle R^2_{\rm rel} \rangle
\end{eqnarray}
where $(Z_1,A_1) = (4,8)$ and $(Z_2,A_2) = (2,4)$ for the $^8$Be + $\alpha$ system.
To apply this formula for charge radii we need to supplement the wave 
functions generated above with a description of the ground state of 
$^8$Be (or at least, of its mean square charge radius). An excellent
description of $\alpha-\alpha$ scattering phase shifts and 
$^8$Be has already been given within the local potential cluster model 
\cite{[BFW]} using a Gaussian nuclear potential
\begin{equation}
V_N(R) = V_G\exp{(-\alpha R^2)} 
\end{equation}
with $V_G=122.6225$ MeV and $\alpha=0.22$  fm$^{-2}$
and a Coulomb potential
\begin{equation}
V_C(R) = {Z_1Z_2 {\rm erf}(\beta R) \over R } 
\end{equation}
with $\beta = 0.75 \ {\rm fm}^{-1}$ and
where $Z_1=Z_2=2$ for the $\alpha-\alpha$ system. We adopt this description
of $^8$Be wholesale because, within our model, it can hardly be improved upon.

As a consistency check on the widths of the states we can make use of our
earlier work (see for example \cite{[pdecay]}) on charged particle decay widths. 
Within the two-body local potential model, a semiclassical approximation 
leads to an $\alpha$-partial width of
\begin{equation}
\Gamma_{\alpha} = F{\hbar^2 \over  4\mu}
\exp{ \left (-2\int_{R_2}^{R_3} dR\ k(R) \right ) },
\end{equation}
where $R_2$ and $R_3$ are the two outermost classical turning points and $\mu$
is the reduced mass of the system. The normalization factor $F$ is given 
to good accuracy by
\begin{equation}
F\int_{R_1}^{R_2} {dR \over 2k(R)} = 1,
\end{equation}
with $R_1$ the innermost turning point, and the wave number $k(R)$ is 
\begin{equation}
k(R) = \left( {2\mu \over \hbar^2} \vert Q-V(R) \vert \right )^{1/2}
\end{equation}
and $Q$ is the experimental energy of the decaying state relative to the 
two-body breakup threshold.

\section{Energies, widths and radii of Hoyle band states}

As outlined in the previous section, we calculate $^8$Be + $\alpha$ cluster
states, using a previously published prescription for the potential \cite{[univ]},
without adjusting any of the parameters $v_0$, $a$ and $x$ listed in
Eq.(3).  We identify the $0^+$
state with the Hoyle state, check that this choice produces a good account 
of the available data, and proceed to calculate similar states of higher angular
momenta. Each band of states is labelled by the choice of $G=2n+L$. 
The maximum possible value of $L$ in this scheme is equal to $G$ itself
(corresponding to the nodeless wave function). We determine the potential
radius $R_0$ of the potential by fitting so as to reproduce the experimental 
energy of the $0^+$ Hoyle state exactly. The resulting values for $R_0$ are listed 
in Table I. It is also useful to know the position and maximum height of the 
potential (Coulomb barrier) since this gives an upper limit on the energy 
of a resonant state with the corresponding value of $L$. Therefore we also 
list these values in Table I for the three values of $G = 4, 6$ and 8 
under consideration.

\begin{table}[htb]
\begin{center}
\caption{\label{tab:table1}
Potential Maxima for $G = 4, 6$ and 8}
\begin{tabular}{cccc}
\hline
\hline
& & &  \\
& \multicolumn{3}{c}{$V_{\rm max}$ (MeV) at $R_{\rm max}  $(fm)} \\
L & $G=4$  &  $G=6$ & $G=8$ \\
& $R_0=0.8883$ fm & $R_0=1.9386$ fm & $R_0=2.9008$ fm \\
& & &  \\
\hline
0 & 1.63 at 6.20 & 1.42 at 7.25 & 1.26 at 8.28 \\
2 & 3.12 at 5.16 & 2.44 at 6.41 & 2.02 at 7.56 \\
4 & 9.23 at 3.48 & 5.74 at 5.28 & 4.24 at 6.62 \\
6 &             & 13.87 at 4.02 & 8.82 at 5.74 \\
8 & &                          & 17.26 at 4.86 \\
\hline
\hline
\end{tabular}
\end{center}
\end{table}

Figure 1 shows the potentials (nuclear, Coulomb and centrifugal combined)
obtained from the procedure described above for $G = 4, 6$ and 8. It is
clear from inspection that it would be no surprise to discover that the 
state with the highest expected $L$-value for each value of $G$ was 
completely unbound (or at best precariously resonant). In fact, none of
the states of the Hoyle band is bound. They are all (at best) resonances, 
and in some cases rather close to the top of the Coulomb barrier, so the 
calculation of their energies and widths is numerically delicate. 
In view of this we have found it expedient
to cross check our results using a variety of different calculational methods. 
\begin{itemize}
\item We use our own bound state code to perform a numerical
integration of the Schr\"odinger equation, but rounding off the
potential at its maximum value for radii in excess of that radius where
the maximum is achieved (i.e. $V=V_{\rm max}$ for $R \geq R_{\rm max}$). 
This serves to give a good estimate of the energies which can be used to 
provide a starting energy for the methods described below, and also as a
consistency check on these subsequent results.
\item We use the published code GAMOW \cite{[Pal]}, which employs complex
arithmetic to solve the Schr\"odinger equation and fits Gamow tails to 
the resonant states, so as to evaluate energies and widths of the states. 
We need this code principally for the wave functions it generates which
can be used eventually in the calculation of the mean square charge radii.
\item We use the published elastic scattering code SCAT2 \cite{[Bersillon]}
which allows us to monitor the scattering phase shifts of the 
$^8$Be + $\alpha$ system as a function of the centre of mass (c.m.)
energy, and thereby to identify the peak energies and widths of the
resonant states in the various partial waves.
\item As an order of magnitude check, we also evaluate the 
widths of the states using the semiclassical method discussed in 
the previous section, Eq.(8). Previous experience using this
approach to calculate half-lives for charged particle decay suggests that
it can be expected to agree with the true value to within about a factor 
of two. 
\end{itemize}

The code SCAT2 writes out the scattering matrix elements $\eta_l$ 
in the form $1 - {\rm Re}(\eta_l)$ and Im($\eta_l$), where the
$\eta_l$  are related to the transmission coefficients $T_l$ by
\begin{equation}
T_l = 1 - \vert \eta_l \vert^2.
\end{equation} 
We run the code at 
successively incremented c.m. energies, centred on the values
indicated by our own bound state code and by GAMOW, and monitor
the behaviour of the scattering matrix as the energy increases
through the suspected resonance region. As the energy of a resonant
state is approached from below, the value of $1 - {\rm Re}(\eta_l)$
for the appropriate partial wave $l$ rises from near zero through 1 
towards 2. We take the energy range over which it rises from 0.5 to
1.5 as the width of the resonant state. The value of Im($\eta_l$)
also rises from 0 towards 1, and then falls again to zero. By systematically
working through the energy regions indicated by the bound state code
(and taking successively finer incremental energy grids as necessary), we
obtain the width values reported for the $L>0$ states in Table II. 
The widths for the $0^+$ states were obtained using the semiclassical
approximation described in the previous section, Eqs.(8-10). We note
that the experimental width of the $0^+$ state is $8.5 \pm 1.0$ eV, 
with which the corresponding theoretical value obtained with $G=6$ 
in Table II is in good agreement.

The energies from all these methods are mutually compatible 
and we present the average of them as the peak energy of each resonance 
in Table II. We have added 7.365 MeV to the c.m. scattering energies to obtain 
excitation energies relative to the $^{12}$C ground state. (Note that the $0^+$ 
state's energy of 7.654 MeV was used to fit the potential radius $R_0$ 
for all three values of $G$)

\begin{table}[htb]
\begin{center}
\caption{\label{tab:table2}
Calculated state energies, widths and $\langle R^2 \rangle $ 
for $G = 4, 6$ and 8}
\begin{tabular}{cccc}
\hline
\hline
& & &  \\
& \multicolumn{3}{c}{$ E{\rm (MeV)} \pm \Gamma$} \\
L & $G=4$  &  $G=6$ & $G=8$ \\
& & &   \\
\hline
0 & $7.654 \pm 3$(eV) & $7.654 \pm 6$(eV) & $7.654 \pm 13$(eV) \\
2 & No state & $9.61 \pm 360$(keV)   & $8.85 \pm 68$(keV)  \\
4 & No state & $13.71 \pm 1.24$(MeV) & $11.52 \pm 300$(keV) \\
6 &             & No state           & $16.25 \pm 660$(keV) \\
8 & &                                & No state \\
& & &   \\
$\langle r^2 \rangle$ & 11.688 {\rm fm}$^2$ & 
13.457 {\rm fm}$^2$ & 15.553 {\rm fm}$^2$ \\
$\sqrt{\langle r^2 \rangle}$ & 3.42 {\rm fm} & 
3.67 {\rm fm} & 3.94 {\rm fm} \\
\hline
\hline
\end{tabular}
\end{center}
\end{table}

Figure 2 shows the energy dependence of $1 - {\rm Re}(\eta_l)$ 
and Im($\eta_l$) for the $G=6, L=2$ resonance near the c.m. energy of
about 2.22 MeV. This is both a typical case, and the state that we are
most interested in (because of recent experimental 
results \cite{[Itoh],[Freer],[Gai]} which have at last found a
rotational state of the Hoyle band). The best description of
this state within our model is achieved using $G=6$. For $G=4$
there is no resonant state at all, and for $G=8$ the excitation 
energy is too low and the width too narrow. We are not unduly concerned
that this value of $G$ does not tally ``nicely" with oscillator 
shell model considerations, which might suggest G=$8$ since most 
of the low-lying intruder states in this region are described as
4p-4h excitations in the shell model. We are not using a harmonic
oscillator potential and will generate rather different wave functions with
exponential rather than Gaussian tails, and so do not expect an 
identity of $G$ values between the two cases. Our calculations also 
predict an excited $4^+$ state at 13.71 MeV. We predict a width of 
1.24 MeV for it, which would make it hard to detect in experiments.
We note however that this excitation energy places the state above the
barrier maximum for an $L=4$ state in a $G=6$ band, indicating that
the result is not completely reliable, although still indicative of a
possible $4^+$ state in this region.

In this context it is interesting to note that Freer et
al.\cite{[Freer4]} have recently found evidence for a new state in
$^{12}$C at $13.3 \pm 0.2$ MeV with a width of $1.7 \pm 0.2$ MeV in
their studies of the $^{12}$C($^4$He, $^4$He+$^4$He+$^4$He) $^4$He and
$^9$Be($^4$He, $^4$He+$^4$He+$^4$He)n reactions. Analysis of the
angular distributions suggests that the state might have
J$^{\pi}=4^+$. As such, its properties are in line with our
expectations for the third state in the Hoyle band, and we await the
outcome of further investigations with interest.

Previous analyses of inelastic electron scattering from $^{12}$C \cite{[Funaki]}
indicate that the Hoyle state has an abnormally large charge radius.
We therefore calculate the values of $\langle r^2 \rangle$ implied
by our model for $G = 4, 6$ and 8. We first need to calculate a mean
square charge radius for $^8$Be. We do this by using the mean square
charge radius for an $\alpha$-particle of 1.6757 fm \cite{[Angeli]} 
and the mean square separation of the two $\alpha$-particles 
calculated by our bound state code and confirmed by GAMOW using 
the potential of Ref.\cite{[BFW]} in the formula
\begin{equation}
\langle R^2 \rangle_{^8{\rm Be}} = \langle R^2 \rangle_{\alpha} + {1 \over 4}
\langle R^2 \rangle_{\alpha-\alpha}
\end{equation}
This yields a mean square charge radius for $^8$Be of 10.634 fm$^2$
(with a square root of 3.261 fm). This, in turn, serves as input for
the mean square charge radius of the Hoyle state in the formula
\begin{eqnarray}
\langle R^2 \rangle_{^{12}{\rm C-Hoyle}} &=& {1 \over 3} \langle R^2 \rangle_{\alpha} + 
{2 \over 3}\langle R^2 \rangle_{{\rm Be}}  \nonumber \\
&+& {2 \over 9}\langle R^2 \rangle_{\alpha-{\rm Be}}
\end{eqnarray}
Our $G=6$ value of 3.78 fm is somewhat below the value of 3.87 fm
deduced by Funaki et al. \cite{[Funaki]} but significantly larger
than the values deduced from earlier GCM calculations of 
3.50 fm \cite{[Uegaki]} and 3.47 fm \cite{[Kamimura]} obtained from 
full 3$\alpha$ calculations using Volkov effective two-nucleon forces. 

\section{Conclusions}

It has long been suspected that the Hoyle state in $^{12}$C
might have rotational excitations built upon it so that a Hoyle band
could be present in the $^{12}$C spectrum. Recent experimental work has located
a $2^+$ state a little below 10 MeV with a width of about 600 keV
which is a strong candidate for such a structure. This has motivated us
to apply a local potential $^8$Be-$\alpha$ cluster model, with previously
published potential prescription, to the system to see if we can
reproduce the existing data and predict properties of other similar 
states.

The calculation is numerically delicate but obtaining closely similar
excitation energies from three different methods give us a good degree
of confidence in our results. We find that with a global quantum number
of $G=6$ we are able to give a good account of the width and root mean 
square charge radius of the Hoyle state itself and a reasonable description 
of the excitation energy and width of the proposed excited $2^+$ member of the
putative Hoyle band. We note that our calculated $2^+$ state is very
close to the top of the Coulomb barrier. We also predict a rather wide 
$4^+$ state of the $G=6$ band at an excitation energy of roughly 14 MeV, and find that 
the band certainly terminates here (if not already with the $2^+$ state). 
However, our calculation is not completely reliable for this state, and there may
be a case to be made that the known $4^+$ state at 14.083 MeV
\cite{[C12data]} should
be assigned to the Hoyle band. Perhaps there are even two $4^+$ states
there. This is an interesting conundrum which further experimental
investigation might be able to resolve.

\begin{widetext}

\begin{figure*}
\vspace{0mm}
\hspace{0mm}
\includegraphics{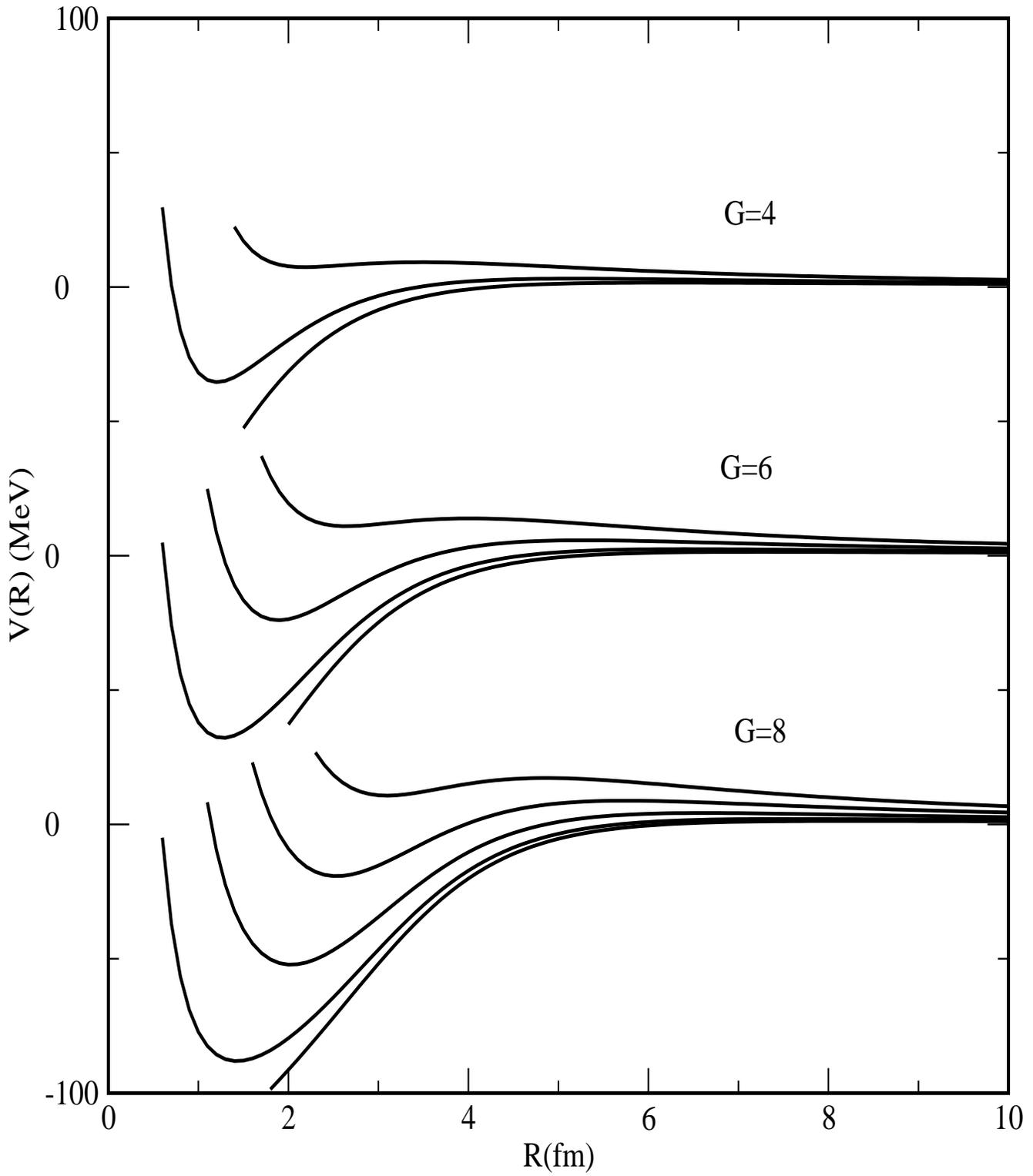}
\vspace{10mm}
\caption{\label{fig1:wide} 
Local potentials (nuclear + Coulomb + centrifugal) for the 
$^8$Be + $\alpha$ system corresponding to all possible $L$-values
for the global quantum numbers $G=4, 6$ and 8. The potentials for $G=4$
are displaced upwards by 100 MeV and those for $G=8$ downwards by 
100 MeV to aid visibility.} 
\vspace{0mm}
\end{figure*}
\end{widetext}

\begin{widetext}

\begin{figure*}
\vspace{0mm}
\hspace{0mm}
\includegraphics{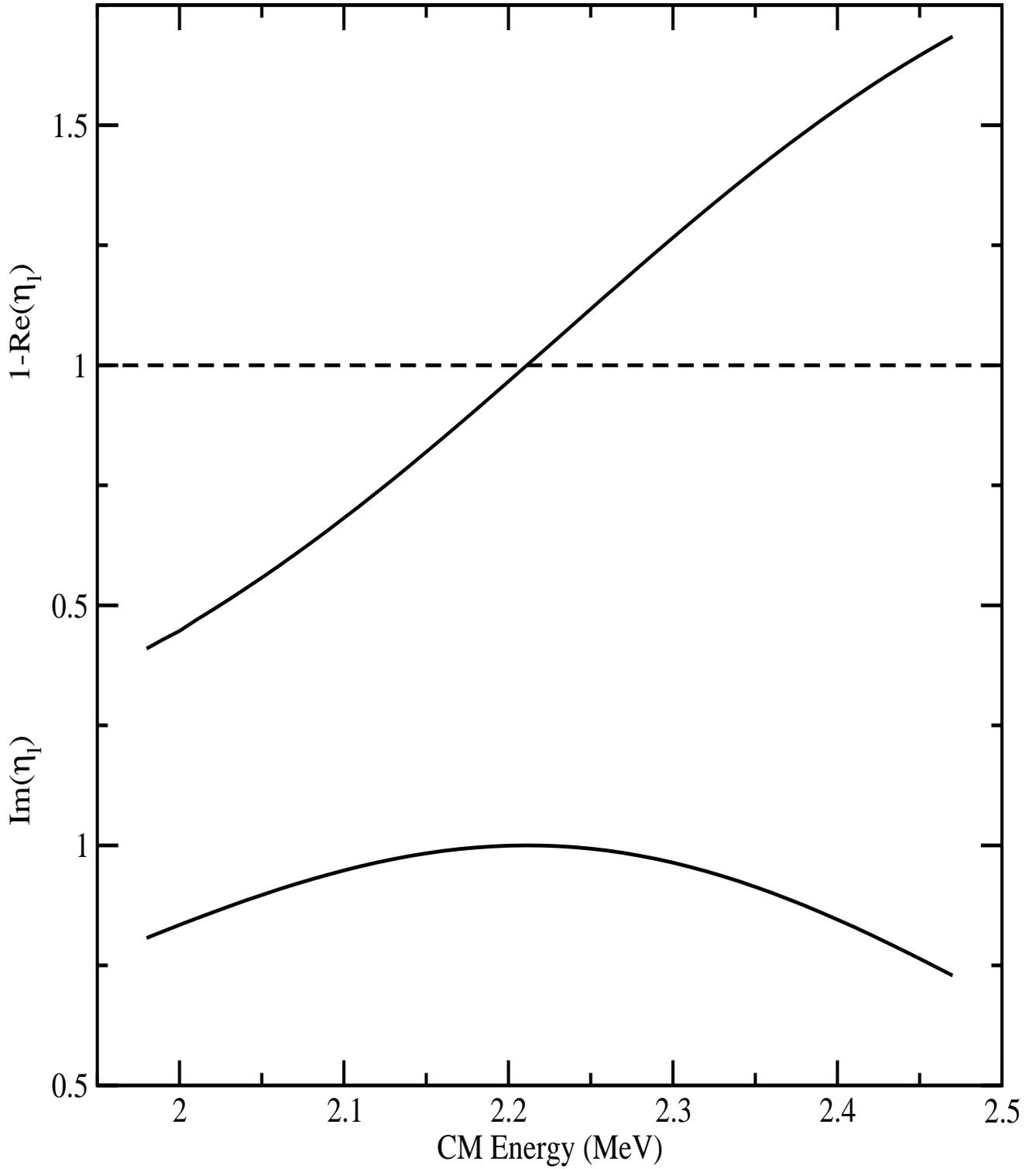}
\vspace{10mm}
\caption{\label{fig2:wide} 
The energy dependence of $1 - {\rm Re}(\eta_l)$ 
and Im($\eta_l$) for the $G=6, L=2$ resonance near the c.m. energy of
2.22 MeV obtained from the code SCAT2. The energy scale has been
shifted slightly to align it with the average energy calculated for
the $2^+$ excitation by our different calculations.} 
\vspace{0mm}
\end{figure*}

\end{widetext}

\end{document}